# Determination of Speed and Source Height of Coronal Shock Waves Using Type II Solar Radio Bursts

G. L. S. S. Liyanage[1], J. Adassuriya[1], K. P. S. C. Jayaratne[1], C. Monstien[2]

[1]*Astronomy and Space Science Unit, Department of Physics, Faculty of Science, University of Colombo, Sri Lanka*
[2]*Istituto ricerche solari Aldo e Cele Daccò (IRSOL), Faculty of Informatics, Università della Svizzera italiana (USI), CH-6605 Locarno, Switzerland.*

*sahanslst@gmail.com*

## 1. ABSTRACT

This study examines the shock speed and source height of coronal shock waves using Type II solar radio bursts. The solar radio burst data from January 2022 to October 2023 were obtained from e-CALLISTO archive. The type II radio bursts were isolated from the spectra through a rigorous noise reduction process by taking the maximum intensity of each time channel. Using plasma oscillations and electron density models for the solar corona, explicit expressions for shock speed and source height were obtained. From the dynamic spectra, the starting frequency of a burst was obtained and using these parameters shock speed and source height were calculated. Results confirmed shock speeds ranging 343-1032 km/s with average speed of $650 \pm 226 \, km/s$ and source heights $1.317 - 1.724 \, R_\odot$, with high precision in formula predictions. The study highlights the need for broader burst type inclusion in future research and underscores the efficacy of the developed methodologies to improve space weather forecasts.

## 2. INTRODUCTION

The Sun profoundly influences the solar system, particularly Earth, through its dynamic processes, including nuclear fusion. This fusion process not only transforms hydrogen into helium but also releases substantial energy that ionizes solar gases, creating strong magnetic fields. These fields, in turn, store vast amounts of energy and often undergo magnetic reconnection a phenomenon where oppositely directed magnetic field lines break and reconnect, converting magnetic energy into kinetic and thermal energy. During this process, the sudden release of energy accelerates charged particles, such as electrons and protons, to near-relativistic speeds, propelling them outward into the solar system in events known as Coronal Mass Ejections (CMEs) and Solar Flares. These energetic particles, driven by reconnection-generated electric fields and turbulence, play a crucial role in space weather, impacting planetary magnetospheres and atmospheric dynamics [1], [2], [3].

Research into these solar phenomena has historically utilized various empirical models to estimate the electron densities within the solar corona, which are critical for understanding the propagation and impact of shock waves driven by solar eruptions. Among these models the Newkirk, Leblanc, and Saito models have each contributed to this understanding by describing these densities with increasing complexity and accuracy[4], [5], [6].

This study focuses on enhancing measurement techniques for solar phenomena, specifically through the analysis of Type II solar radio bursts (SRBs). These bursts, essential for determining the shock speed and source height of coronal shock waves, offer a direct method to probe the mechanics of solar eruptions and enhance the prediction accuracy of related space weather events. By leveraging the global e-Callisto network, this research analyzes metric-type Type II SRBs to refine the methodologies used in space weather forecasting.

Utilizing the Newkirk density model, this paper seeks to derive precise formulas for shock speed and source height, comparing these with data obtained through traditional methods. The goal is to improve the accuracy of predictions regarding the timing and impact of solar activities on Earth's technological infrastructure and space environment[7], [8].

This paper contributes to the field by proposing advanced methods for data preprocessing and analysis, which include novel approaches for noise reduction and burst isolation in solar radio data. These enhancements are aimed at bolstering the reliability of measurements obtained from Type II SRBs, thus supporting more accurate forecasts of space weather conditions and mitigating the adverse effects of solar activities.

## 3. THEORY

### 3.1 Newkirk Density Model

The Newkirk density model is a semi-empirical model designed to describe the electron density in the solar corona. The model is especially relevant for studies related to solar radio bursts, such as Type II and Type III bursts, which influenced by the density of the corona where they propagate. Assuming the spherical symmetric corona, the Newkirk density model is represented by,

$$n_e(R) = n_0 \times 10^{\alpha \frac{R_\odot}{R}} \qquad (1)$$

Where $n_0 = 4.2 \times 10^{-4} \ cm^{-3}$ is the electron density near at the base of the corona, $R_\odot$ is the solar radius, $R$ is the radial distance measured from the center of the sun to an arbitrary point in the solar corona and $\alpha = 4.32$ [4], [5], [6].

### 3.2 Plasma Frequency

Plasma frequency is a fundamental property of plasma. The plasma frequency represents the natural oscillation rate of the electron density in a plasma in response to a small perturbation without the presence of an external electric field[9]. The expression for the plasma frequency is given by,

$$f_{pe} = \frac{1}{2\pi}\sqrt{\frac{n_e e^2}{\epsilon_0 m_e}} \qquad (2)$$

Substituting constants for the above equation and then we can approximate the plasma frequency as follows:

$$f_{pe} \approx 8.978 \times 10^{-3}\sqrt{n_e}\ MHz \qquad (3)$$

### 3.3 Derivation of the expressions for Source Height and the Shock Speed

Using the electron density and plasma frequency from the above electron density and the plasma frequency by combining both equations it is possible to derive an expression for the shock height ($R_s$):

$$R_s = \frac{\alpha R_\odot \ln(10)}{\ln\left(\frac{f_{pe}^2}{n_0 \kappa^2}\right)} \qquad (4)$$

Where $R_s$ is the distance of the shock relative to the center of the sun, $R_\odot$ is the radius of the sun, $\alpha = 4.32$, $n_0 = 4.2 \times 10^{-3}\ cm^{-3}$, and $\kappa = 8.978 \times 10^{-3}\ MHz/cm^{\frac{3}{2}}$.

By getting the time derivative of the $R_s$, an expression for the Shock speed can be obtained as follows:

$$V_s = \frac{2R_\odot \alpha \ln(10)}{f_{pe} \ln^2\left(\frac{f_{pe}^2}{n_0 \kappa^2}\right)} \cdot \left|\frac{df_{pe}}{dt}\right|\ km/s \qquad (5)$$

Substituting corresponding constant values, the shock speed can be estimated from:

$$V_s = \frac{1.385 \times 10^7}{f_{pe} \ln^2\left(0.295 f_{pe}^2\right)} \cdot \left|\frac{df_{pe}}{dt}\right|\ km/s \qquad (6)$$

Where $\left|\frac{df_{pe}}{dt}\right|$ is the frequency drift rate which can be directly obtained from the Type II dynamic spectrum.

## 4. METHODOLOGY

### 4.1 Data Collection

The data were collected by the e-CALLISTO solar radio spectrometer network[10], publicly available at e-CALLISTO database. 25 well-defined Type II solar radio bursts were selected. The observation dates for bursts were between January 2022 and October 2023.[1]

## 4.2 Data Preprocessing

### 4.2.1 FITS Reading

The downloaded data files from the e-CALLISTO database are in the "fit.gz" format which is known as the compressed FITS file format. By employing Python 3.10 with 'Astropy' library the 'fit.gz' file was opened and read. A FITS file contains two headers: the first header stores the numerical values of intensity data for a radio signal detected by the e-Callisto network as an $n \times m$ matrix, while the second header contains the 'frequency' and 'time' data for a radio signal. Utilizing this information, the complete picture of radio signals was plotted using the 'matplotlib' library. As mentioned, the data consists of an $n \times m$ matrix, where $n$ represents the frequency channel, and $m$ represents the time channel. Consequently, the matrix values indicate the intensities of the radio signal for corresponding frequency and time channels (Figure 1).

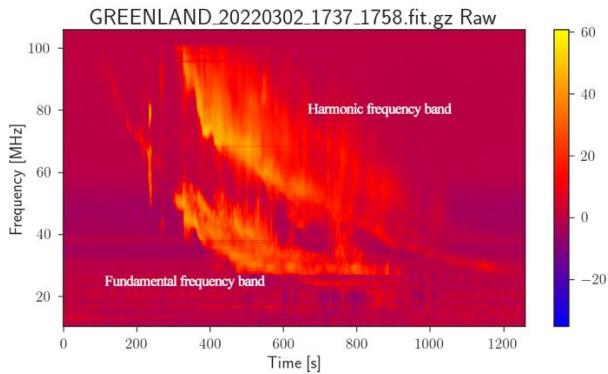

Figure 1 The raw plot of a FITS file.

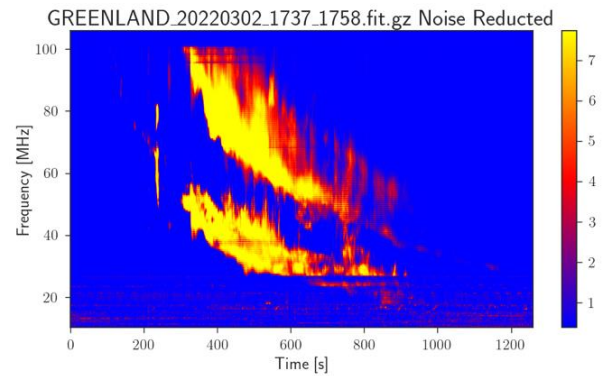

Figure 2 Noise reduced FITS file.

---

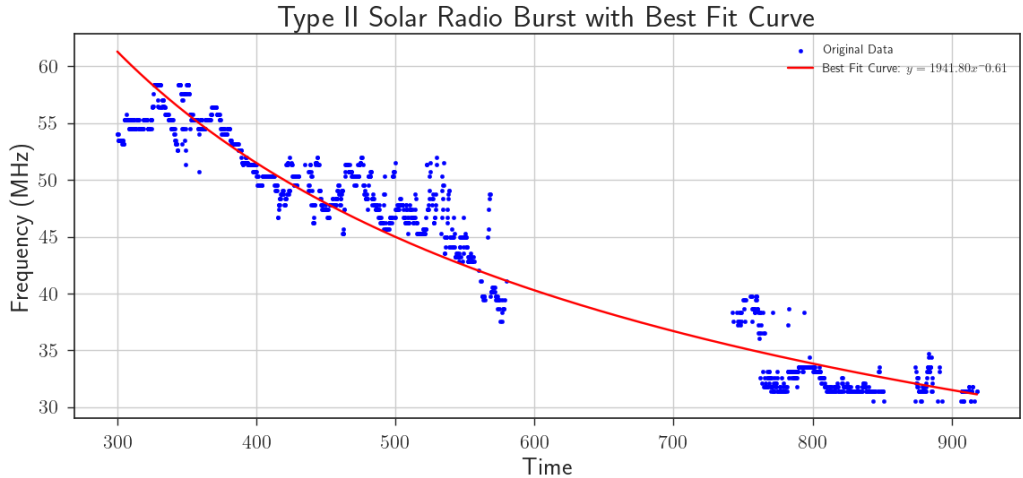

Figure 3 Isolated burst and fitted curve for isolated burst.

### 4.2.2 Background Noise Reduction

The numerical observation data extracted from the FITS file comprises an $n \times m$ matrix. The mean $(\mu_i)$ of intensity values of each row is subtracted from every intensity data point in the matrix, resulting in the following filtered data:

$$data_{ij} = data_{ij} - \mu_i$$

Subsequently, the filtered data was further processed using a filter that sets threshold values.

$$f(x) = \begin{cases} -L, & if\ data_{ij} < 0 \\ data_{ij}, & if\ L < data_{ij} < U \\ U, & if\ U < data_{ij} \end{cases}$$

For each numerical value in the data, the lower threshold $L$ and the upper threshold $U$ are defined uniquely for each dataset. If a data value falls below the lower threshold, it is replaced with $L$; if it exceeds the upper threshold, it is replaced with $U$. Otherwise, the data value remains unchanged. The final data values are normalized and plotted (Figure 2).

### 4.2.3 Burst Isolation

Isolating the fundamental frequency band is essential to calculate the shock speed. This was accomplished by loading the filtered data into Python and then manually setting the margins of the data matrix to select the fundamental band. Subsequently, all other data values were set to zero, effectively isolating the burst's fundamental band. Then the maximum intensity for each frequency channel was obtained and plotted against the burst time. Even though the burst is isolated from the FITS image, it still contains some noise from the harmonic band. To remove these outliers, an area containing the outliers was selected using the two horizontal lines and two vertical lines, and the outliers were removed. Several manual filters were applied to obtain the isolated fundamental band of the Type II radio burst (Figure 3).

### 4.2.3 Calculation of Shock Speed and Source Height

To calculate the shock speed and the source height, the obtained filtered dataset was fitted to a power law function in the form of $f = A \times t^b$. In this paper we are interested in finding the initial shock speed it is necessary to find an estimation for the starting frequency of the burst. The average value of the highest 90[th] percentile of all frequencies was chosen as the starting frequency of the burst. Using the starting frequency and the frequency drift rate from the graph, the source height and the shock speed was calculated (Figure 3).

## 5. RESULTS AND DISCUSSION

### 5.1 Validation of the Noise Reduction Procedure

In this study, all chosen Type II radio bursts were metric type, which typically encompasses solar radio bursts within a wavelength range of 0 to 10 meters. The obtained frequency drift rates were plotted against the starting frequencies of the corresponding bursts.

It was observed that the frequency drift rate $\left(\frac{df_{pe}}{dt}\right)$ and starting frequency ($f_s$) to a power law function as $\frac{df_{pe}}{dt} = a \times f_s^{\psi}$. As per Aguilar et al. (2005), for metric-type, Type II solar radio bursts, the power law index ($\psi$) stands at 1.44 [11]. In this study, the power law index ($\psi$) was determined to be $1.45 \pm 0.42$, suggesting that the noise reduction procedure employed is suitable for further analysis. Additionally, the correlation coefficient indicated a reasonable association between the frequency drift rate and the starting frequency. Although the accuracy of the power law index is deemed satisfactory, the precision of the measurements raises concerns. Thus, it underscores the need for enhancing the noise reduction process. During the burst isolation process, it was noted that some significant intensity points of the burst were lost, which resulted in less precise measurements, as evidenced by the larger variance.

## 5.2 Analysis of Shock Speed

The average shock speed for the Type II bursts was determined to be $650 \pm 226 \ km/s$, with the calculated shock speed spanning from $343 \ km/s$ to $1032 \ km/s$. These findings agree with previous studies[12].

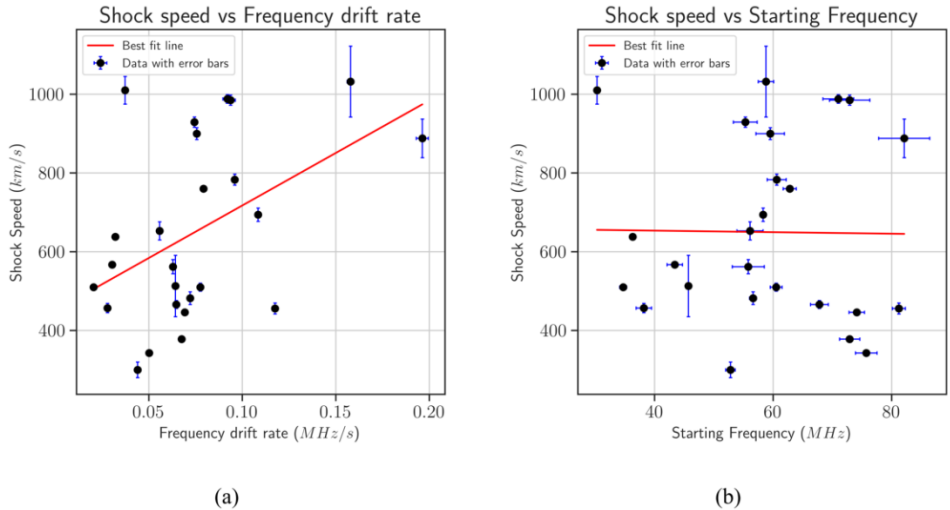

(a)                                           (b)

Figure 4 The graphs of shock speed vs frequency drift rate and shock speed vs starting frequency

The Figure 4a illustrates the relationship between shock speed and frequency drift rate. The correlation coefficient of 0.463 suggests a weakly positive correlation between shock speed and drift rate, which agrees with the expectations set forth by Equation 1; the equation posits that shock speed is directly proportional to the absolute value of the drift rate. Conversely, Figure 4b depicts the variation of shock speed with the starting frequency, where a correlation coefficient of $-0.012$ indicates an absence of correlation between the two variables. This outcome is consistent with predictions from Equation 1, which states that shock speed does not have a direct or inverse proportional relationship with the starting frequency. Furthermore, it is evident that the error in calculating the shock speed increases with higher shock speeds. The starting frequency was determined using the percentile method, which also exhibits increasing errors as the starting frequency increases

## 5.2 Analysis of Source Height

The source height of the coronal shock, also known as the shock formation height, represents the initial radial distance from the center of the Sun. It was determined to be in the range of $1.317 R_{\odot}$ to $1.724 R_{\odot}$, consistent with existing literature.

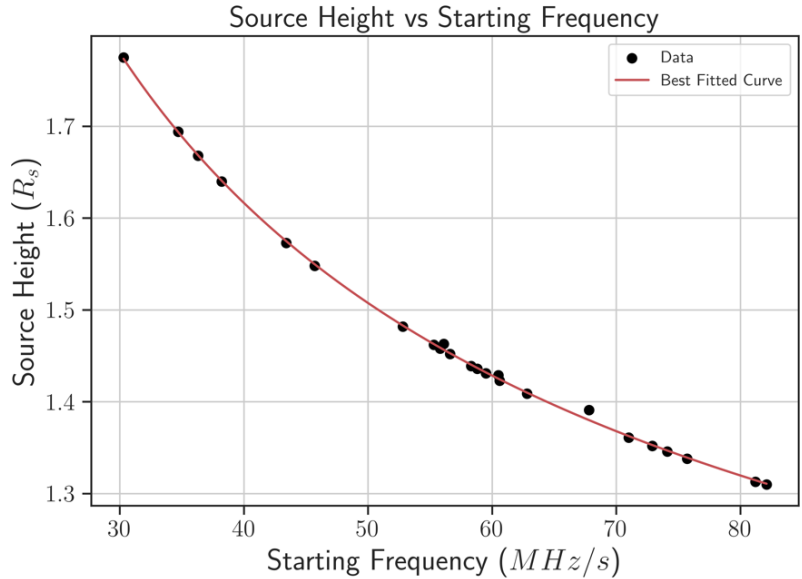

Figure 5 The graph of source height vs starting frequency.

The best-fit curve for the graph of source height $(R_s)$ vs starting frequency $(f_s)$ is represented by the equation $R_s = 7.7 f_s^{-0.59} + 0.73$ (Figure 5). The robustness of the model is supported by an R-squared value of 0.999 and a Root Mean Square Error (RMSE) of 0.003, indicating an excellent fit to the data points. Within the metric wavelength range of Type II solar radio bursts, the source height $(R_s)$ can be accurately predicted using this equation, provided that the starting frequency $(f_s)$ is known.

## 6. CONCLUSION

This study analyzed 25 datasets of Type II solar radio bursts to calculate shock speed and source height of coronal shock waves, utilizing a Python-based noise reduction process. Results confirmed the symmetric distribution of shock speed (343-1032 km/s) with the average shock speed was given by 650 ± 226 km/s, and source height (1.317-1.724$R_\odot$), frequency drift rate, and starting frequency. Derived formulas from the Newkirk model showed strong agreement with historical data, particularly the relationship $R_s = 7.7 f_s^{0.59} + 0.73$ with an $R^2$ value of 0.999. The power law index for frequency drift rate was 1.45 ± 0.42 validating the analytical methods and enhancing theoretical solar physics and space weather forecasting capabilities.